\documentclass[10pt]{article}
\usepackage[utf8]{inputenc}
\usepackage{enumerate}
\usepackage[utf8]{inputenc}
\usepackage{amsmath}
\usepackage{amsfonts}
\usepackage{amssymb}
\usepackage{graphicx}
\usepackage{color}
\usepackage{mathptmx}
\usepackage{multicol}
\usepackage{float}
\usepackage{subcaption}
\RequirePackage[numbers,sort&compress]{natbib}
\RequirePackage[colorlinks=true,citecolor=blue,urlcolor=blue,linkcolor=blue]{hyperref}
\usepackage[left=2cm,right=2cm,top=2cm,bottom=2cm]{geometry}

\author{Tanima Duary\footnote{Email: td14ip021@iiserkol.ac.in}, \quad Ananda Dasgupta\footnote{Email: adg@iiserkol.ac.in} \quad and \quad Narayan Banerjee\footnote{Email: narayan@iiserkol.ac.in} \\ {\normalsize Department of Physical Sciences, 
Indian Institute of Science Education and Research Kolkata,
Mohanpur - 741 246, WB, India}}
\title{{\Large\bf Thawing and Freezing Quintessence Models: A thermodynamic Consideration}}
\begin{document}
\date{}
\maketitle
\vspace{-0.5cm}
\begin{abstract}
Thawing and freezing quintessence models are compared thermodynamically. Both of them are found to disobey the Generalized Second Law of Thermodynamics. However, for freezing models, there is still a scope as this breakdown occurs in the past, deep inside the radiation dominated era, when a standard scalar field model with a pressureless matter is not a correct description of the matter content. The thawing model has a pathological breakdown in terms of thermodynamics in a finite future.
\end{abstract}

\smallskip
\quad\textbf{Keywords}: Thawing models, freezing models, generalized second law 

\quad\textbf{PACS}: 98.80.-k; 98,89.Jk
\medskip

\section{Introduction}

Convincing observational evidences of the accelerated expansion of the universe\cite{perl, schmidt, riess, knopp, planck} for the last twenty years posed the great challenge of modelling the universe or more precisely finding a matter component that can bring such a repulsive gravity effect into being. The natural choice, the cosmological constant $\Lambda$ fails to be the unique choice for its unassailable discrepancy between the actually required and theoretically predicted values\cite{paddy}. A scalar field with a potential, called the quintessence field, is one of the favoured options although no potential warrants a compelling theoretical support\cite{brax, sami, varun}. \\

In connection with their evolution pattern, some of the quintessence fields are broadly classified into two categories called thawing and freezing models. The thawing model is one for which the effective equation of state parameter $w$ starts as almost a constant close to $-1$ and ``thaws" into an evolving one, whereas a freezing model behaves in a different manner, $w$ settles down to a constant value close to $-1$ quite late in the evolution of the universe. Scherrer and Sen\cite{anjan} provides a brief but elegant description of this classification. Particularly interesting amongst the freezing models are the so called ``trackers''\cite{stein1, stein2, johri}, in which the energy density of the scalar field evolves almost parallel to the energy density of the dark matter for the most of the history, without dominating over the latter but freezes to a value more than the corresponding dark matter density at a later stage of the evolution.  As such a quintessence field does not necessarily have to belong to one of these two categories, but both the thawing and freezing models are important in their own right and attracted a lot of attention, primarily by virtue of their ability to address the coincidence issue\cite{concord} - the question why the dark matter and dark energy are of the same order of magnitude now. \\

A comparison between freezing models with a tracking behaviour and a thawing model, in connection with their compliance with observational data has been given by Thakur, Nautiyal, Sen and Seshadri\cite{shruti}. These models are also compared in connection with their stability\cite{abhijit}. It is found that there is hardly any reason to favour any of these two if the complete history of universe in the post radiation-dominated is considered. Thawing and freezing models, in terms of cluster number counts, have been discussed by Devi, Gonzales and Alcaniz\cite{Devi2014}. The purpose of the present work is to compare the thawing and freezing models considering their thermodynamic behaviour, particularly their viability against the generalized second law (GSL) of thermodynamics. A very general description of thawing and freezing models are adhered to, and tracking behaviour or any such finer details are not considered. \\

We use a simple definition of thawing and freezing models and plot the rate of change of the total entropy against evolution. This rate should always be positive as the total entropy never decreases according to the GSL. The results show that both forms suffer from a breakdown of GSL. We have tested this for a quintessence along with a cold dark matter (CDM), but also do the same exercise with a pure quintessence both leading to similar results to indicate that this feature of thermodynamic non-compliance is a characteristic of the quintessence field. The freezing models, however, has an advantage over the thawing models as this breakdown of GSL occurs way back in the past ($z\sim 10^{4}$), which is not really described by a CDM with a quintessence. \\

The next section contains a general description of the quintessence models. Section 3 describes a pure quintessence without any dark matter content and describes the thermodynamic behaviour of the thawing and freezing models. Section 4 deals with the thermodynamics of such models along with the dark matter component. The fifth and final chapter contains a discussion of the results obtained.

\section{Quintessence Models: Thawing and freezing}

The model of the universe is considered to be filled with a perfect fluid and a scalar field $\Phi$ endowed with a potential $V(\Phi)$. The relevant action is 

\begin{equation} \label{action}
S^{field} = \int \mathrm{d}^4 x \sqrt{-g}\Big[ \frac{R}{16\pi G}-\frac{g^{\mu\nu}}{2} \mathrm{\partial} _\mu \Phi \mathrm{\partial} _\nu \Phi-V(\Phi)+\mathcal{L}_m\Big],
\end{equation} 

where $R$ is the Ricci scalar and $\mathcal{L}_m$ is the Lagrangian density for the fluid distribution. In the consequent Einstein field equations $G_{\alpha\beta} = - T^{(f)} _{\alpha\beta} - T^{(q)} _{\alpha\beta}$, where the superscripts $f$ and $q$ represent the fluid and the quintessence matter respectively, the right hand side is given by,

\begin{equation} \label{em-f}
T^{\alpha (f)} _\beta = (\rho+p)u_\alpha u_\beta+pg_{\alpha\beta},
\end{equation}
and
\begin{equation}\label{em-q}
T_{\alpha\beta}^{(q)} = \Big[\mathrm{\partial} _\alpha \Phi \mathrm{\partial} _\beta \Phi-\frac{1}{2}g_{\alpha\beta}g^{\mu\nu} \mathrm{\partial} _\mu \Phi \mathrm{\partial} _\nu \Phi - 2g_{\alpha\beta} V(\Phi)\Big].
\end{equation} 
Here $\rho, p$ represent the density and pressure of the fluid respectively and $u_\alpha$ is the fluid velocity vector and is normalized as $u^\alpha u_\alpha=-1$. For a comoving observer, one can write $u^\alpha=\delta^\alpha_0$. \\

The universe is assumed to be isotropic, homogeneous and spatially flat, given by the Friedmann-Robertson-Walker metric,
\begin{equation} \label{metric}
\mathrm{d}s^2 = -\mathrm{d}t^2+{a(t)}^2 [\mathrm{d}r^2+r^2 \mathrm{d}\theta^2+r^2\sin^2\theta \mathrm{d}\phi^2],
\end{equation}
where $a(t)$ is the scale factor. In such a spacetime $\Phi$ is also a function of the cosmic time $t$ alone. Einstein field equations can be written as
\begin{equation}\label{fe1}
 3 H^2 = 3 \Big(\frac{\dot{a}}{a}\Big)^2 = 8\pi G (\rho + \rho_\Phi),
\end{equation}

\begin{equation}\label{fe2}
 2 \frac{\ddot{a}}{a} + \Big(\frac{\dot{a}}{a}\Big)^2 = - 8\pi G (p + p_\Phi),
\end{equation}
where $H$ is the Hubble parameter, $\rho_\Phi, p_\Phi$ are the density and pressure of the quintessence field given by

\begin{equation}\label{q-den}
\rho_\Phi \equiv T^0_0= \frac{1}{2}(\dot\Phi)^2+V(\Phi)
\end{equation} 
and
\begin{equation}\label{q-press}
p_\Phi \equiv T^1_1= T^2_2=T^3_3=\frac{1}{2}(\dot\Phi)^2-V(\Phi).
\end{equation}

The fluid conservation equation

\begin{equation}\label{mat-cons}
 \dot{\rho} + 3H(\rho + p) = 0.
\end{equation}

The Klein-Gordon equation for $\Phi$ is

\begin{equation}\label{kg-phi}
\ddot{\Phi}+3H\dot{\Phi}+V^\prime(\Phi)=0,
\end{equation} 

which is not an independent equation in view of the field equation and equation (\ref{mat-cons}). In the following two sections, we shall discuss the thermodynamic viability of the pure scalar model and more realistic model containing both the scalar field and a pressureless dust. \\

\section{A pure quintessence}

To begin with, a model without a cosmic fluid is considered. So in equations (\ref{fe1}) and (\ref{fe2}), $p=\rho=0$. One now has two equations to solve for three unknowns, $a, \phi, V$, as equation (\ref{kg-phi}) is still not an independent equation. To close the system of equations, we pick up the ansatz for $\rho_\Phi$ as given by Carvalho {\it et al}\cite{Carvalho2006},

\begin{equation} \label{rhophi-ansatz}
\frac{1}{\rho_\Phi}\frac{\mathrm{\partial}\rho_\Phi}{\mathrm{\partial} a} = -\frac{\lambda}{a^{1-2\alpha}},
\end{equation}
where $\lambda(>0)$ and $\alpha$ are parameters. This readily integrates into 

 \begin{equation}\label{rhophi}
  \rho_\Phi(a)= {\rho_{\Phi,0}} \exp\left[-\frac{\lambda}{2\alpha}(a^{2\alpha}-1)\right],
 \end{equation}
where  a subscript $0$ indicates the present value of the quantity and $a_0$, the present value of scale factor is taken to be $1$. In the limit $\alpha\rightarrow 0 $ the quintessence energy density reduces to a power law, $\rho_\Phi(a)\propto a^{-\lambda}$. Using equation \eqref{kg-phi}, the potential $V$ is obtained as (see \cite{Carvalho2006} for the details),

\begin{equation}\label{potential}
V(\Phi) = \left[1-\frac{\lambda}{6}(1+\alpha\sqrt{\sigma}\Phi)^2\right]\rho_{\Phi,0}\exp\left[-\lambda\sqrt{\sigma}(\Phi+\frac{\alpha\sqrt{\sigma}\Phi^2}{2})\right].
\end{equation}
As $\lambda$ was chosen to be positive so that the energy density of the quintessence field decreases with the evolution, $\sigma = \frac{8\pi G}{\lambda}$ is positive and $V$ is a real valued function of $\Phi$. In the limit $\alpha\rightarrow0$, the above equation reproduces the exponential potential. \\

The equation of state parameter (EoS) for the scalar-field, $w_\Phi \equiv \frac{p_\Phi}{\rho_\Phi}$. In this model, takes the form-
\begin{equation}\label{14}
w_\Phi=-1+\frac{\lambda}{3}a^{2\alpha}.
\end{equation}
 The above EoS is time dependent quantity (for $ \alpha\neq 0$) through the scale factor $a$. In the limit $\alpha \rightarrow0$, it reduces to a constant, $w_\Phi = -1+\frac{\lambda}{3}$. \\

\begin{figure}
\centering     
\begin{subfigure}[]{0.45\textwidth}\boxed{\includegraphics[width=76.5mm]{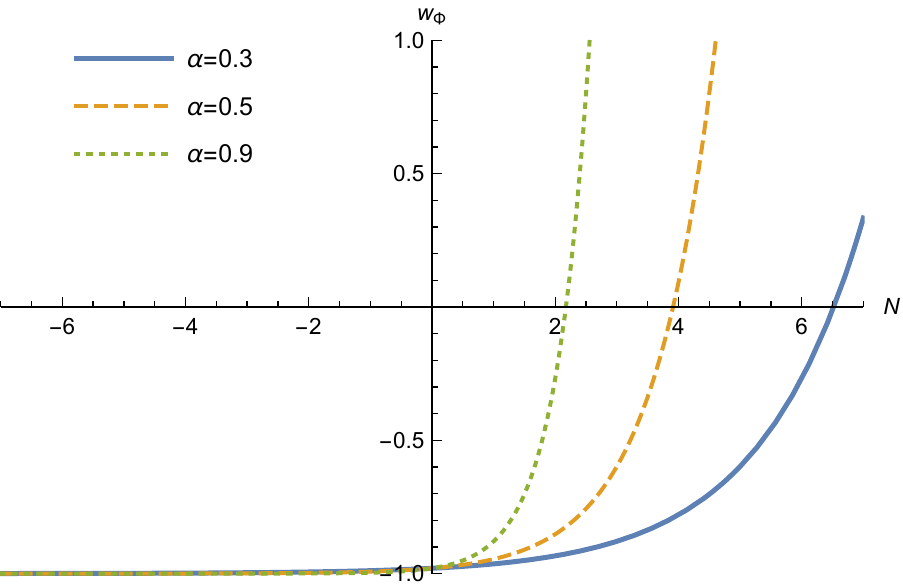}}
\caption{}
\label{fig1a}
\end{subfigure}
\begin{subfigure}[]{0.45\textwidth}\boxed{\includegraphics[width=76.5mm]{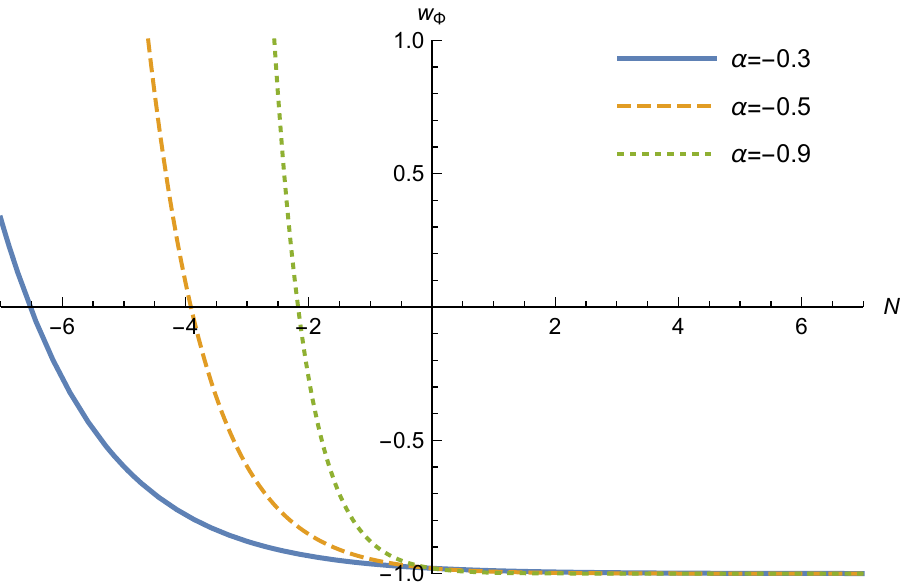}}
\caption{}
\label{fig1b}
\end{subfigure}
\caption{$w_\Phi$ is plotted as a function of $N$. (a) For positive values of $\alpha$, EoS increases from $~-1$, i.e., it exhibits a thawing behavior. (b) For negative values of $\alpha$, EoS $>-1$ decreases to more negative value , i.e., it exhibits a freezing behavior.}
\end{figure}

In the fig-\ref{fig1a} and \ref{fig1b},  $w_\Phi$ is plotted as a function of $N$ for some values of $\alpha$, where $N=\ln(\frac{a}{a_0})$. For positive and negative values of $\alpha$, one has a thawing [Fig. \ref{fig1a}] and a freezing [Fig. \ref{fig1b}]  behaviour respectively \cite{Devi2014}.  \\ 

One may note that, since $\lambda$ is positive, $w_\Phi\geqslant -1$ for all values of $a$, no  matter whether $\alpha$ is positive or negative \cite{Carvalho2006}. \\

The equation (\ref{rhophi-ansatz}) is a general ansatz. The classification into thawing and freezing is now taken care of by the parameter $\alpha$. If $\alpha > 0$, the scalar field has a thawing behaviour while if $\alpha < 0$, it has a freezing behaviour. In the first case the equation of state parameter $w_{\Phi}$ starts from a almost flat value close to $-1$ in the past and melts into lesser negative values, whereas second one decreases to more and more negative values to attain a plateau close to $-1$ in the future. We should clearly mention that the ansatz (\ref{rhophi-ansatz}) has been borrowed from the work Carvalho {\it et al}\cite{Carvalho2006}. The values of the parameters $\alpha, \lambda$ etc. are also taken from reference \cite{Devi2014}.
 
Substituting $\rho_\Phi$ from equation \eqref{rhophi} in the Friedmann equation \eqref{fe1} with $\rho =0$ one obtains the solution for the scale factor as, 

\begin{equation}\label{solution-1}
\frac{1}{2\alpha}\big[\Gamma(0,-\frac{\lambda}{2\alpha})-\Gamma(0,-\frac{\lambda}{4\alpha}a^{2\alpha}) \big] =\gamma (t-t_0),
\end{equation}

where $\Gamma(a,x)$ is the upper incomplete gamma function and is defined as, $\Gamma(a,x)=\int^\infty_x z^{a-1}\exp(-z)dz$. And $\gamma=\sqrt{\frac{8\pi G }{3}\rho_{\Phi,0}}\exp\big(\frac{\lambda}{4\alpha}\big)$ is a constant term. The solution is quite involved but still can be used to check the thermodynamic viability of the model. 

\subsection{Compliance of a pure quintessence with GSL} 

We now investigate whether a pure quintessence model comply with the Generalized Second Law (GSL) of thermodynamics which asserts that the change in the combination of the horizon entropy and entropy of the matter inside the horizon does
not decrease with time\cite{Bekenstein1973, Bekenstein1974}. \\

The total entropy is sum of horizon entropy ($S_h$) and the matter (inside the horizon) entropy ($S_{in}$) i.e.,
\begin{equation} \label{totalentropy}
S_{tot}=S_h+S_{in}.
\end{equation}
For an evolving universe it is prudent to consider the entropy of dynamical apparent horizon rather than teleological event horizon. Entropy of the apparent horizon is given by the relation \cite{Bak2000,Ferreira2016, Hawking1975},
\begin{equation}
S_h=\frac{A}{4G},
\end{equation}
where $A$ denotes the area of the apparent horizon and is related to radius of the apparent horizon ($R_h$) as, $A = 4\pi R_h^2$. In a spatially flat FRW space, the horizon radius $R_h$ is related to Hubble parameter \cite{Bak2000,Ferreira2016, book:Faraoni} as,
\begin{equation}\label{hor-rad}
R_h = \frac{1}{H}.
\end{equation}

Therefore, rate of change in horizon entropy is 
\begin{align} \label{hor-entropy}
\dot{S}_h =-\frac{2\pi}{G}\Big(\frac{\dot{H}}{H^3}\Big).
\end{align}

Applying Gibb's law of thermodynamics for the matter (the quintessence field in this case) inside the horizon one can write, 
\begin{equation} \label{gibbs}
T_{in}\mathrm{d}S_{in} = \mathrm{d}E_{in}+p_\Phi \mathrm{d}V_h.
\end{equation}
The rate of change in entropy of the matter inside the horizon can be written as,
\begin{equation}\label{s-matter-dot}
\dot{S}_{in}=\frac{1}{T_{in}} [(\rho_\Phi + p_\Phi )\dot{V}_h+\dot{\rho}_\Phi V_h],
\end{equation}
where the volume $V_h=\frac{4}{3}\pi R_h^3$. If we consider that the quintessence matter is in thermal equilibrium with the horizon, then $T_{in}$ is same as the temperature of the dynamical apparent horizon($T_h$) i.e., Hayward-Kodama temperature \cite{book:Faraoni, Helou2015,Rani2018,Criscienzo2007, Cai2005}, which is written as,
\begin{equation} \label{hor-temp}
T_{h} = \frac{2H^2+\dot{H}}{4\pi H}.
\end{equation}

It is easily seen that for a de Sitter space(for which $\dot{H}=0$), this temperature reduces to the Hawking temperature, $T_{\text{Hawking}}=\frac{H}{2\pi}$ \cite{Hawking1974}. \\

The rate of change of entropy inside the horizon can be written using eqation (\ref{s-matter-dot})is obtained as (see \cite{Debnath2012}),

\begin{align}\label{s-matter-dot2}
\dot{S}_{in} = \frac{(\rho_\Phi+p_\Phi)}{T_{in}}4\pi R_h^2\big[\dot{R_h}-HR_h\big],
\end{align}
which can be simplified to, 
\begin{equation}\label{s-matter-dot3}
\dot{S}_{in} =\frac{2\pi}{G}\Big(\frac{\dot{H}}{H^3}\Big)\Big(1+\frac{\dot{H}}{2H^2+\dot{H}}\Big).
\end{equation}

Using equations \eqref{hor-entropy} and \eqref{s-matter-dot3}, we get the rate of change of the total entropy as,
\begin{equation}\label{s-tot-dot}
\dot{S}_{tot} = \dot{S}_h + \dot{S}_{in} = \frac{2\pi}{G}\Big(\frac{\dot{H}^2}{H^3}\Big)\Big(\frac{1}{2H^2+\dot{H}}\Big).
\end{equation}
Using the solution (\ref{solution-1}) one can write $H$ and $\dot{H}$ in terms of $a$ and the equation (\ref{s-tot-dot}) looks like 
\begin{equation}\label{s-tot-dot1}
\dot{S}_{tot} =\frac{\pi \lambda^2}{G}\Big(\frac{1}{\sqrt{\frac{8\pi G}{3}\rho_{\Phi,0}}\exp\big(-\frac{\lambda}{4\alpha}(a^{2\alpha}-1)\big)}\Big)\Big(\frac{a^{4\alpha}}{4-\lambda a^{2\alpha}}\Big).
\end{equation}

To check the thermodynamic viability of the model we now plot $\dot{S}_{tot}$ against $N$.

\begin{figure}[h!]
\centering     
 \begin{subfigure}[]{0.75\textwidth}\boxed{\includegraphics[width=125mm]{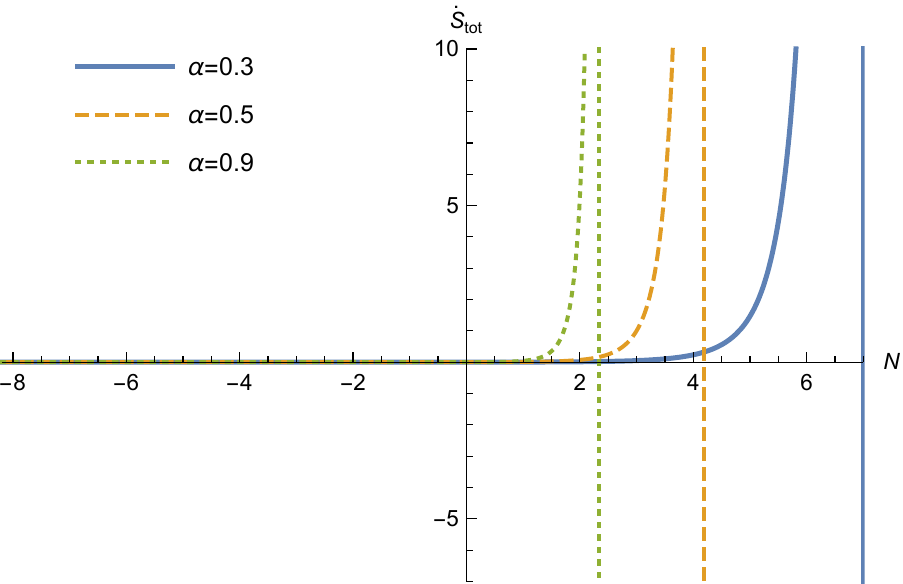}}
 \caption{}
 \label{fig2a}
 \end{subfigure}
\begin{subfigure}[]{0.75\textwidth}\boxed{\includegraphics[width=125mm]{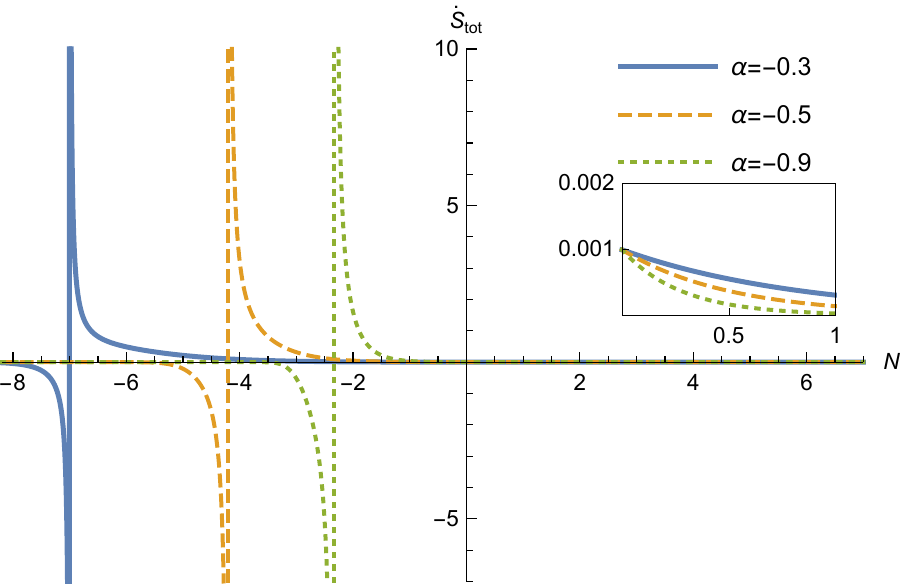}}
\caption{}
\label{fig2b}
\end{subfigure}
\caption{ $\dot{S}_{tot}$ is plotted as a function of $N$. (a) in thawing scenario (for  $\alpha=0.3, 0.5$ and $0.9$) (b) in freezing scenario (for $\alpha=-0.3, -0.5$ and $-0.9$ )}
\end{figure}

From fig-\ref{fig2a}, it is evident that in case of thawing quintessence ($\alpha > 0$), the total entropy increases upto a certain time, after that it does not obey GSL. In fact $\dot{S_{tot}}$ shoots to an infinitely large value and then drops to an infinitely large negative value. The same feature is obtained for all allowed values of $\alpha$, only the range of $N$ for the onset of this pathological behaviour varies. The freezing quintessence ($\alpha < 0$) behaves in an opposite way! It does qualify this inquest for the future, the net entropy increases and settles down to a constant value in future when $\dot{S_{tot}}$ approaches zero (see fig-\ref{fig2b}), may mot be as fast as shown in the figure but rather asymptotically as revealed by the zoomed in version in the inset. However,it does not respect GSL in the past. Here also one has a discontinuity in $\dot{S_{tot}}$ and also a negative value for the same indicating a decrease in $S_{tot}$.  \\

The value of $\lambda$, which determines the rate of fall of the dark energy energy density $\rho_\Phi$ (see equation \ref{rhophi-ansatz}) has a small postive value\cite{Devi2014}. In all the figures $\lambda = 0.06$ is used. We have checked with substantially lower ($\lambda = 0.01$) and higher ($\lambda = 0.1$) values as well. As there is hardly any difference in the qualitative features, we excluded them in order to avoid repetitions.

\section{Quintessence with cold dark matter}

In this section, a more realistic model is considered, where a pressureless fluid (baryonic matter and cold dark matter) is present along with the quintessence matter. The stress-energy component due to dust is given by equation (\ref{em-f}) with $p=0$. So the field equations (\ref{fe1} , \ref{fe2}) will have an input $p=0$. As the Klein-Gordon equation is not an independent equation, we have two equations, (\ref{fe1} , \ref{fe2}), to solve for three quantities, $a, \Phi, V$. We take the same ansatz \eqref{rhophi-ansatz} as in the previous section to close the system of equations.  A direct integration of equation (\ref{mat-cons}) leads to, 
 \begin{equation}\label{rho-mat}
 \rho =\frac{\rho_{0}}{a^3},
 \end{equation}
where $\rho_{0}$ is energy density of matter at present epoch. \\
  
The deceleration parameter $q=-\frac{\dot{H}+H^2}{H^2}$ takes the form, 
 \begin{align} \label{dec-par}
q(a)= \frac{3(-1+\frac{\lambda}{3}a^{2\alpha})+1+\frac{\frac{\rho_{m,0}}{a^3}}{{\rho_{\Phi,0}} \exp\big[-\frac{\lambda}{2\alpha}(a^{2\alpha}-1)\big]}}{2(1+\frac{\frac{\rho_{m,0}}{a^3}}{{\rho_{\Phi,0}} \exp\big[-\frac{\lambda}{2\alpha}(a^{2\alpha}-1)\big]})}.
\end{align}

\begin{figure}[h!]
\centering     
\begin{subfigure}[]{0.45\textwidth}\boxed{\includegraphics[width=76.5mm]{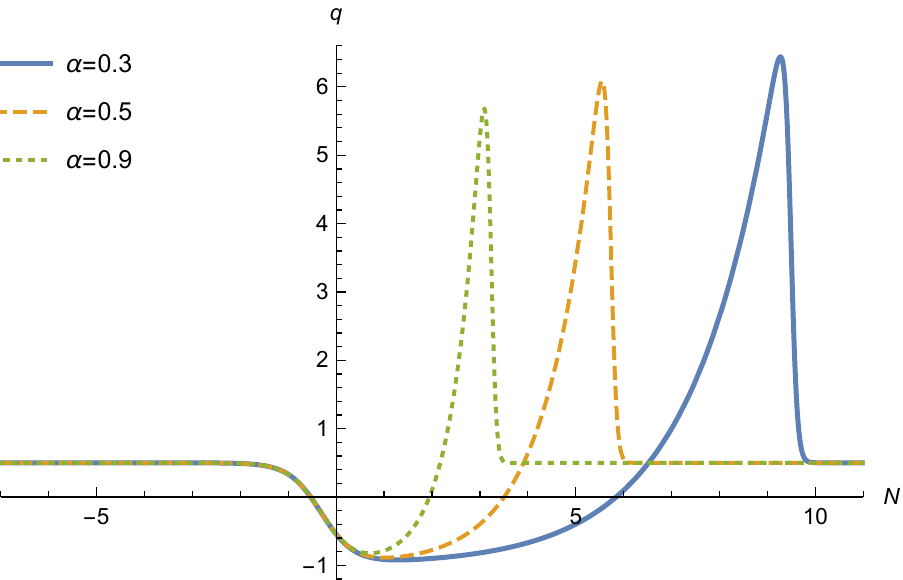}}
\caption{}
\label{fig3a}
\end{subfigure}
\begin{subfigure}[]{0.45\textwidth}\boxed{\includegraphics[width=76.5mm]{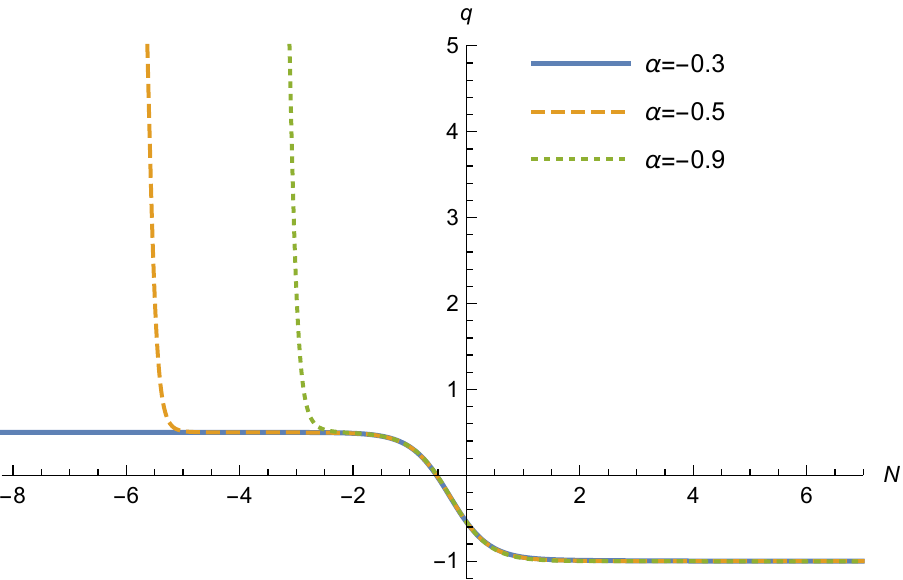}}
\caption{}
\label{fig3b}
\end{subfigure}
\caption{$q$ is plotted as a function of $N$ . (a)  In Thawing scenario , transient acceleration is inevitable ,(b) Freezing quintessence leads the universe into eternal accelerating phase. }
\end{figure}

Fig-\ref{fig3a} brings out the interesting possibility that the present accelerated expansion is a transient phenomenon. In a finite future the universe will re-enter he decelerated phase. This feature was already noticed, such as by Carvalho {\it et al}\cite{Carvalho2006} and Devi {\it et al}\cite{Devi2014}. The freezing model, on the other hand, settles down to a universe which accelerates its expansion once it transits out from the decelerated phase.

These features, as already mentioned, are already known in the literature. The present aim is to check the thermodynamic viability. 

\subsection{Compliance of CDM plus quintessence distribution with GSL} 
  
We consider that the fluid inside the horizon is in thermal equilibrium with the apparent horizon. Hayward-Kodama temperature, as given in equation (\ref{hor-temp}), is used in the following analysis as in the previous section. Now, total energy density is $\rho_{tot}=\rho + \rho_\Phi$. The general form of the rate of change of the total entropy remains same as equation \eqref{s-tot-dot}. As a consequence of modification of the Friedmann equations due the contribution of the cold dark matter, one obtains the explicit form of ${\dot{S}}_{tot}$ as, 
\begin{align}\label{s-tot-dot3}
\dot{S}_{tot} &=\frac{3\sqrt{3\pi}}{2\sqrt{2}G^{3/2}}\frac{{\big[\rho_{\Phi,0}\exp[-\frac{\lambda}{2\alpha}(a^{2\alpha}-1)](\frac{\lambda}{3}a^{2\alpha})+\frac{\rho_{m,0}}{a^3}\big]}^2}{{\big[\rho_{\Phi,0}\exp[-\frac{\lambda}{2\alpha}(a^{2\alpha}-1)]+\frac{\rho_{m,0}}{a^3}\big]}^{3/2}} \nonumber \\
&\times \frac{1}{\big[\rho_{\Phi,0}\exp[-\frac{\lambda}{2\alpha}(a^{2\alpha}-1)](4-\lambda a^{2\alpha})+\frac{\rho_{m,0}}{a^3}\big]}.
\end{align}

\begin{figure}[h!]
\centering     
\begin{subfigure}[]{0.75\textwidth}\boxed{\includegraphics[width=125mm]{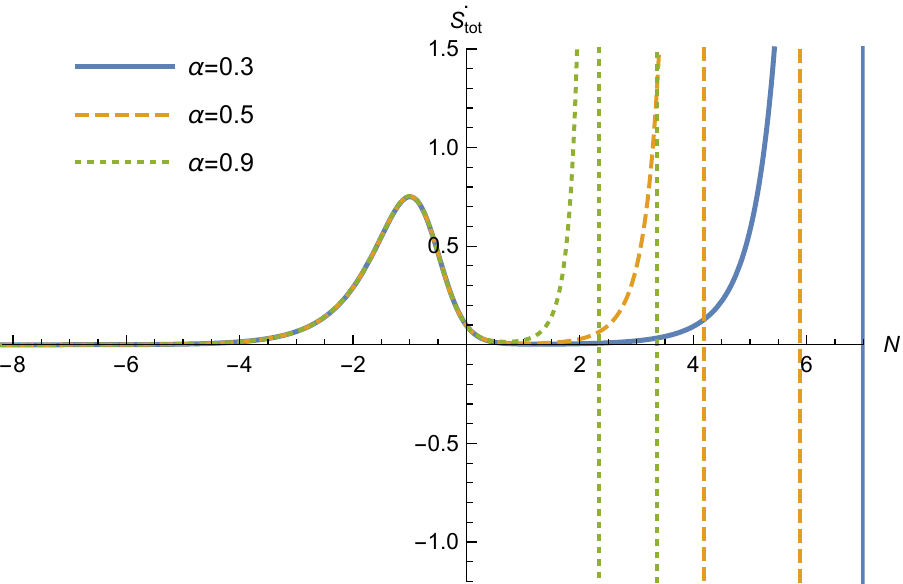}}
\caption{}
\label{fig4a}
\end{subfigure}
\begin{subfigure}[]{0.75\textwidth}\boxed{\includegraphics[width=125mm]{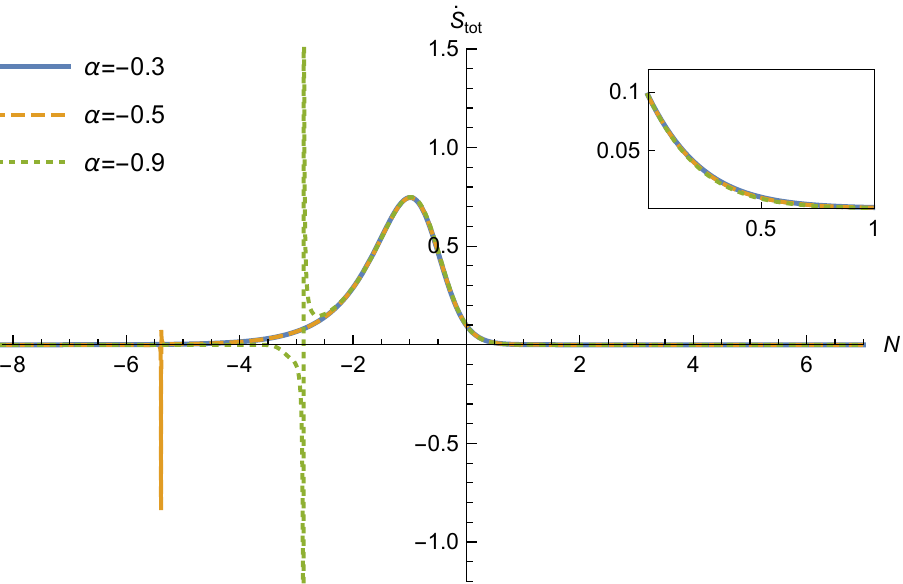}}
\caption{}
\label{fig4b}
\end{subfigure}
\caption{$\dot{S}_{tot}$ is plotted as a function of $N$ . (a) in thawing scenario (for  $\alpha=0.3 , 0.5$ and $0.9$) (b) in freezing scenario (for $\alpha=-0.3 , -0.5$ and $-0.9$ ) }
\end{figure}

\begin{figure}[h!]
\centering     
\begin{subfigure}[]{0.45\textwidth}\boxed{\includegraphics[width=76.5 mm]{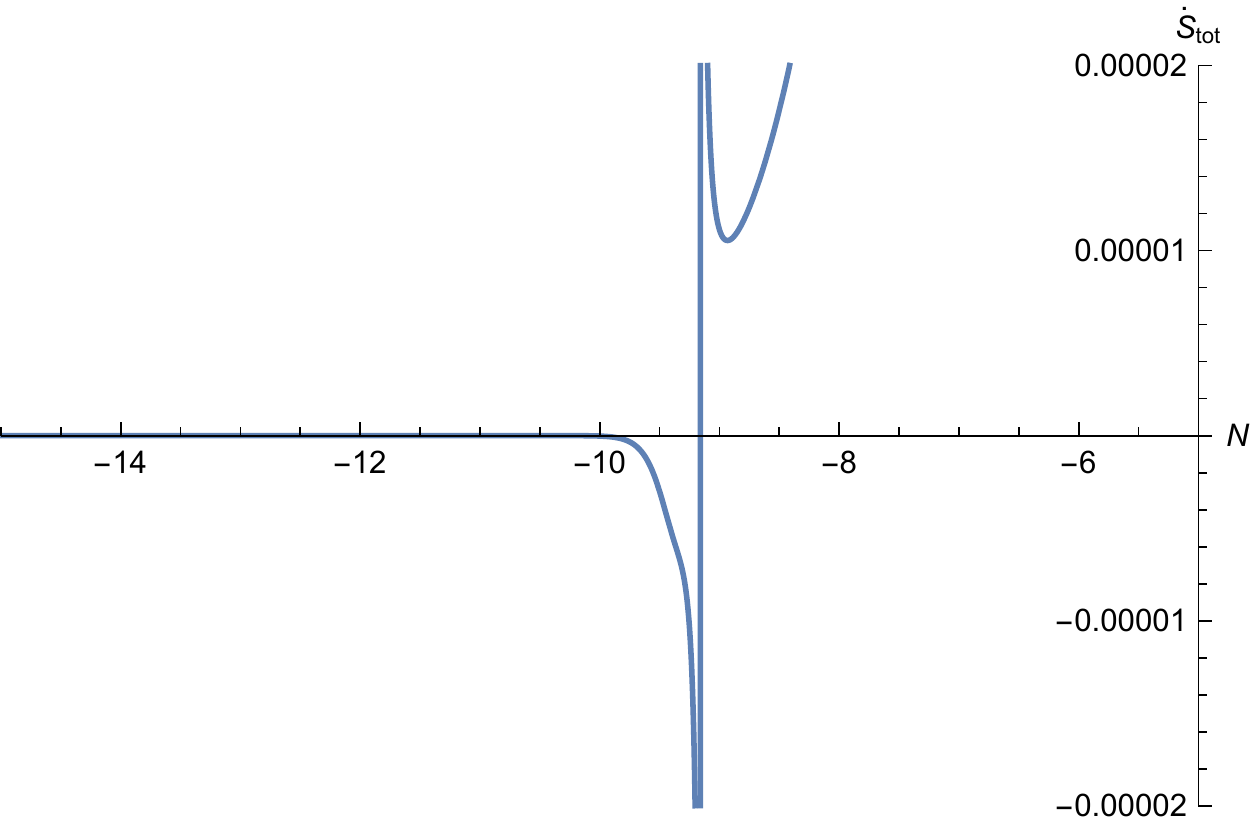}}
\caption{}
\label{fig5a}
\end{subfigure}
\begin{subfigure}[]{0.45\textwidth}\boxed{\includegraphics[width=78.5 mm]{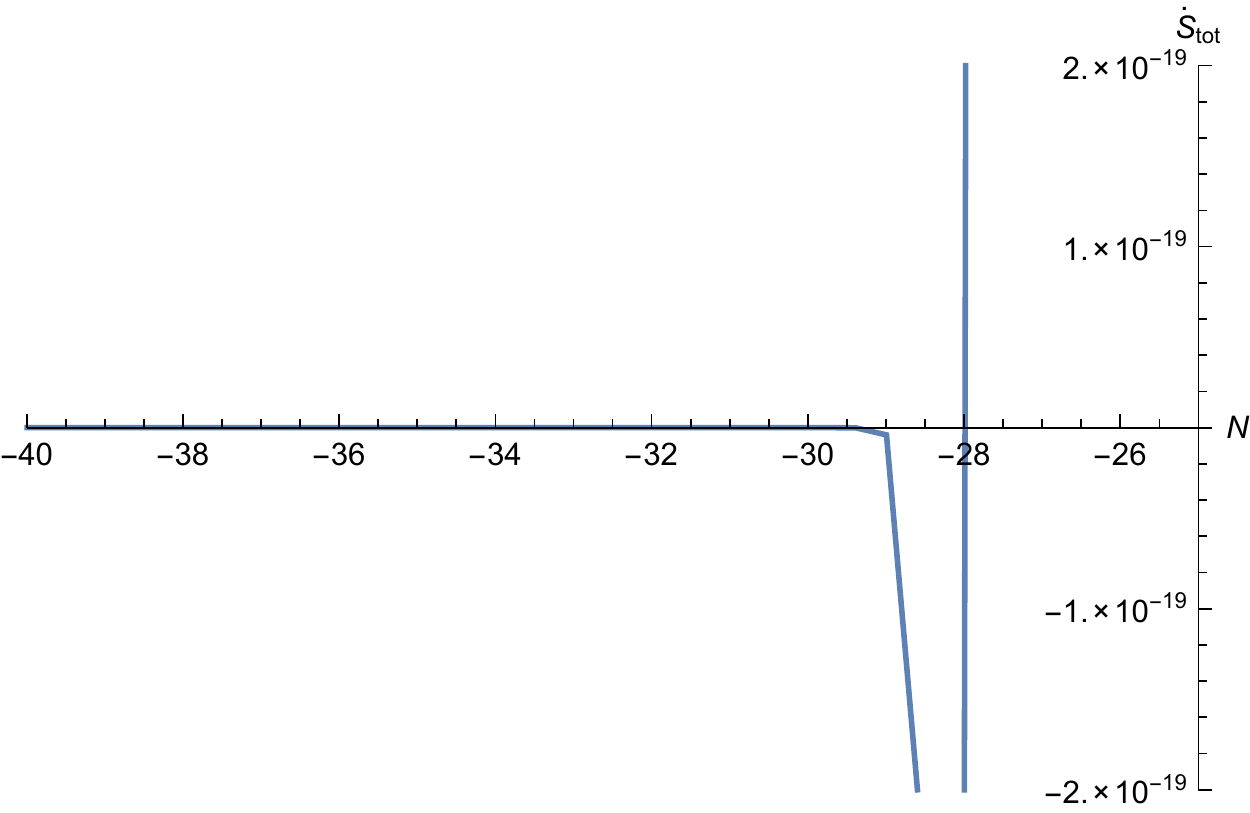}}
\caption{}
\label{fig5b}
\end{subfigure}
\caption{Rate of change of total entropy is plotted as a function of N in Freezing scenario for (a) $\alpha=-0.3$ and (b) $\alpha=-0.1$. }
\end{figure}

The qualitative features remain the same as in the case of a pure quintessence. As evident from figure \ref{fig4a}, the thawing models ($\alpha > 0$) will violate the GSL in some finite future as $\dot{S}_{tot}$ is negative at some finite $N$ indicating a decrease in entropy. The freezing models ($\alpha < 0$) do respect GSL in the future. Albeit having a decreasing value, $\dot{S}_{tot}$ remains positive (as seen in the figure 4(b)), indicating that the entropy is increasing and settles down to a constant value in future, as indicated by figure \ref{fig4b}. In past, however, it indeed has a problem. The rate of change of entropy becomes negative at a finite past. However, by the choice of the parameter $\alpha$, the time of occurrence can be pushed back. For instance, for $\alpha = -0.3$, the pathology is observed for $z \sim  10^{4}$ (see figure \ref{fig5a}), i.e., before the onset of matter domination over the radiation, where this system of equations will not govern the dynamics of the universe. Similarly, for $\alpha=-0.1$, this discrepancy is observed at $z\sim 10^{12}$, which is far beyond the jurisdiction of quintessence along with CDM (see figure \ref{fig5b}). \\

If we carefully notice Eq. \eqref{s-tot-dot}, it is apparent that the term $2H^2+\dot{H}$ decides the fate the thermodynamic viability of the models.  For $\dot{H} + 2H^2 < 0$ (i.e., $q \geq 1$), the model fails in the thermodynamic inquest. So the violation of GSL is associated with a decelerated universe in future. The deceleration has to be at least as rapid as in the case of a pure radiation dominated universe. However, no alarm is indicated for the good old standard radiation dominated model of the universe which does not contain any dominating scalar field - equation (\ref{s-tot-dot3}) clearly indicates that 
if $\rho_{\Phi,0} = 0$, $\dot{S}_{tot}$ is a positive semi-definite quantity. \\

So one can see that the freezing models are stronger against thermodynamic viability, at least in the relevant period when the system under consideration is actually valid. \\

As in the previous section, here also we have used the figure for $\lambda = 0.06$, and excluded $\lambda = 0.01$ and $0.1$ as there is no change in the quality. The only difference is that of a minor shift in the epochs, such as that of the onset of the violation of GSL.


\section{Discussion}

Thawing and freezing models are compared in this work in the context of the Generalized Second Law of thermodynamics. The total entropy ($S_{tot}$) is taken as the sum of the horizon entropy and the entropy of the matter inside the horizon. With a simple ansatz\cite{Devi2014} for the evolution of the energy density of the quintessence field, one can easily figure out the range of the parameter ($\alpha$) values responsible for thawing and freezing behaviour of the field. It is found that both of them are incompatible with GSL, there are situations where $S$ decreases, and decreases very fast. For the freezing models, this breakdown can occur at a distant past ($z \sim 10^{4}$) where a quintessence model along with a CDM does not account for the evolution, one has to have dominant contribution from a radiation distribution. So this breakdown of GSL may not be real. For the thawing models, however, this pathological breakdown of GSL is in a finite future. Thus the major indication is that the freezing models are favoured compared to the thawing ones on the considerations of thermodynamic viability.

\section{Acknowledgment}
Tanima Duary wants to acknowledge CSIR for funding this project. She also wishes to thank her colleagues for lively  discussions.

\end{document}